\documentclass[iop]{emulateapj}
\usepackage{verbatim}
\usepackage{rotating}
\usepackage{graphicx}
\usepackage{natbib}
\usepackage{epsfig}
\usepackage{array}
\usepackage{amsmath}
\usepackage[latin1]{inputenc}
\usepackage{multirow}
\usepackage{xcolor}

\shorttitle{Labeling Bias in Galaxy Morphologies}

\begin{document}
\title{Systematic Labeling Bias in Galaxy Morphologies}
\author{Guillermo Cabrera-Vives}
\affil{Department of Computer Science, University of Concepci\'on, Edmundo Larenas 219, Concepci\'on, Chile}
\affil{Millenium Institute of Astrophysics, Chile}
\email{guillecabrera@udec.cl}
\author{Christopher J. Miller}
\affil{Department of Astronomy and Department of Physics, University of Michigan, 1085 S. University, Ann Arbor, MI 48109, USA}
\author{Jeff Schneider}
\affil{School of Computer Science, Carnegie Mellon University, 5000 Forbes ave, Pittsburgh, PA 15213, USA}

\begin{abstract}
  We present a metric to quantify systematic labeling bias in galaxy morphology data sets stemming from the quality of the labeled data. This labeling bias is independent from labeling errors and requires knowledge about the intrinsic properties of the data with respect to the observed properties. We conduct a relative comparison of label bias for different low redshift galaxy morphology data sets. We show our metric is able to recover previous de-biasing procedures based on redshift as biasing parameter. By using the image resolution instead, we find biases that have not been addressed. We find that the morphologies based on supervised machine-learning trained over features such as colors, shape, and concentration show significantly less bias than morphologies based on expert or citizen-science classifiers. This result holds even when there is underlying bias present in the training sets used in the supervised machine learning process. We use catalog simulations to validate our bias metric, and show how to bin the multidimensional intrinsic and observed galaxy properties used in the bias quantification. Our approach is designed to work on any other labeled multidimensional data sets and the code is publicly available\footnote{https://github.com/guille-c/labeling\_bias}.
\end{abstract}



\section{Introduction}

For more than a century astronomers have been working to understand
galaxy properties and evolution from their morphology. The seminal examples is
the Hubble sequence \citep{Hubble1926}, which first classified galaxies into
ellipticals, spirals, barred spirals, and irregulars.
Galaxy morphologies have been shown to correlate with other intrinsic properties
such as color, brightness, maximum rotation velocity, and
gas content \citep{Dressler1980}. 
From these properties, it is possible to infer important
physical properties such as stellar population fraction, surface star
density, total mass, and gas to star conversion rates \citep[see][and references therein]{Odewahn2002}.

For some time, visual classifications played the dominant role in galaxy
morphologies. Classifications have been done by expert astronomers
\citep{deVauc76, deVauc91, Bundy2005, Fukugita2007, Schawinski2007, Nair2010, Kartaltepe2015} as well as non-expert citizen scientists
through crowdsourcing systems such as Galaxy Zoo \citep{Lintott2008, Bamford2009, Lintott2011, Willett2013, Simmons2017, Willett2017}. With the advent of large high quality survey data like from the Sloan
Digital Sky Survey \citep{York2007} and CANDELS \citep{Grogin2011}, we
are beginning to see more machine learning morphologically classified galaxy data sets using a
variety of methods \citep[e.g.][]{Ball2004, Scarlata2007, Tasca2009, Gauci2010, Huertas2011, Dieleman2015, Huertas2015}. Many of the current machine learning classification techniques fall into the category of supervised learning and thus require training data sets, usually based on visually classified morphologies. Examples of unsupervised machine learning classifications can be found in \cite{Naim1997, Edwards2013, Kramer2013, Shamir2013, Schutter2015}.

When visually classifying galaxies according to their morphologies, the resulting
labels will be biased in terms of observable parameters. Low
resolution and dim galaxy images will be biased towards smoother
types, because the human annotator in charge of labeling the images
will not be able to see the fine structure of these objects. Bias in galaxy morphology catalogs has been extensively studied by the Galaxy Zoo team \citep{Lintott2008}. In \cite{Bamford2009} and \cite {Willett2013} a bias correction term was applied to morphology probabilities by assuming that the morphological fraction does not evolve over the redshift within bins of fixed galaxy physical size and luminosity. For Galaxy Zoo: Hubble morphologies \citep{Willett2017} artificially redshifted images have been used to quantify this bias. A different way of addressing the problem is through a machine learning approach, simultaneously learning a classification model, estimating the intrinsic biases in the ground truth, and providing new de-biased labels \citep{Cabrera2014, Bootkrajang2016}.

In this paper, we present a metric for measuring this labeling bias in
 morphological classification data sets and we compare low redshift morphological catalogs of spiral/elliptical galaxies from experts \citep{Fukugita2007, Nair2010}, non-experts \citep{Lintott2011} and machine learned \citep{Huertas2011}. We release to the public the code for measuring labeling bias and for simulating multi-dimensional labeling bias. This code can be used not only by the galaxy evolution community but also by anyone interested in measuring how biased their catalogs are in terms of observable parameters.
 
This paper is organized as follows: in Section \ref{sec:bias} we develop a statistical measure of labeling bias based on the fraction of objects in terms of their intrinsic and observable properties. Our metric is based upon the assertion that the fractions of labels are fixed within bins on the intrinsic properties. We then quantify variations in labeled fractions from the estimated intrinsic fraction as a function of observed properties. In Section \ref{sec:data} we describe the data sets to be used and how we created simulated galaxy morphology biased data sets. Some considerations on the bias-variance trade-off of our estimators have to be taken in to account. This is explained in Section \ref{sec:MeasuringGalaxyBias}, where we also describe the methodology used to address this issue. In Section \ref{sec:GalaxyBias} we measure the biases for different data sets and show that even ``expert
labels'' are often biased in terms of observed quantities like apparent size. In Section \ref{sec:Conclusions} we describe the main conclusions coming from this work.

\section{Classification Bias}
\label{sec:bias}

In real data it may be very hard to obtain the high quality true
classification labels $y_i$, which we will call the \emph{ground
  truth} or \emph{gold standard}. However, one can always make an
estimate of this ground truth $\hat{y}_i$. In supervised and
semi-supervised machine learning this is usually accomplished through
human annotators. In terms of galaxy morphology, the estimated labels
stem from visual inspection of galaxy images. These visually
determined morphologies are sometimes used directly in scientific
analyses. Sometimes, they are used explicitly to train classification
algorithms (e.g., supervised learning). Sometimes they are used
implicitly to test such algorithms or used in conjunction with
unlabeled data (e.g., semi-supervised learning). However, the galaxies
are always convolved with the point spread function (PSF) of the
telescope, which makes it difficult to visually (or even
computationally) resolve the spiral features in small and faint
galaxies. For galaxies, this means that in the estimated labels, spirals
can be misclassified as ellipticals. This labeling bias is more
important when the PSF is close to the angular size of the galaxies,
particularly for ground based telescope classification, such as Galaxy
Zoo. As noted in \citet{Bamford2009} and \citet{Cabrera2014}, this
bias is not statistical nor inherent to the visual classifiers, but a
direct consequence of the quality of the data.

There are many steps which go into the visual classification of the
morphologies of galaxies. While we expect classifiers to notice that the
light profile is steeper for ellipticals than for disk-like spirals,
classifiers also use color and spatial feature identification
during their classification process. Depending on the filters used or
the resolution of the galaxy image, it is possible to confuse one type
of morphology with another. Worse, these mislabellings can be
consistent amongst different human classifiers leading to a high
degree of statistical confidence in the wrong classification label. 

\begin{figure}
\plottwo{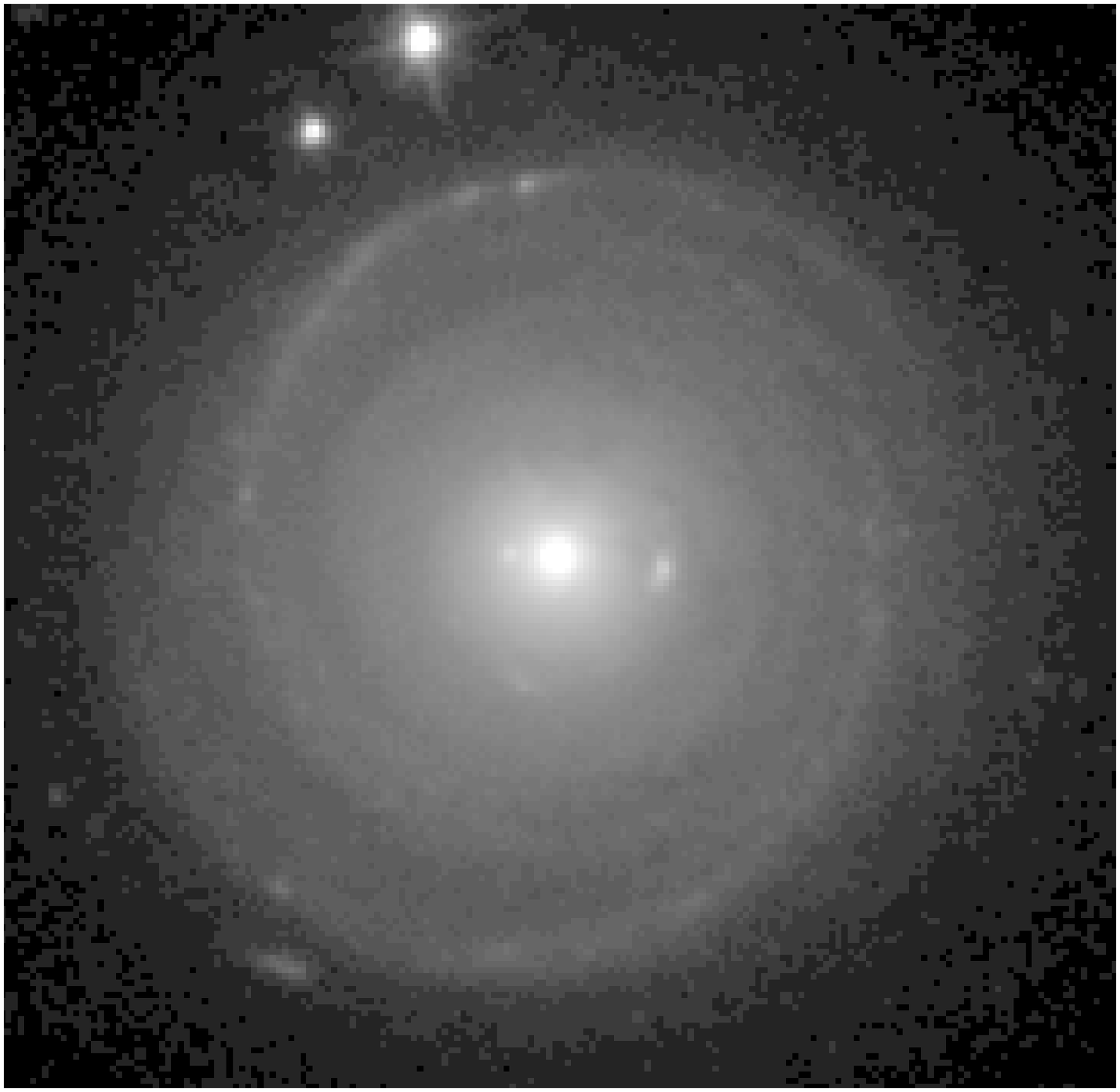}{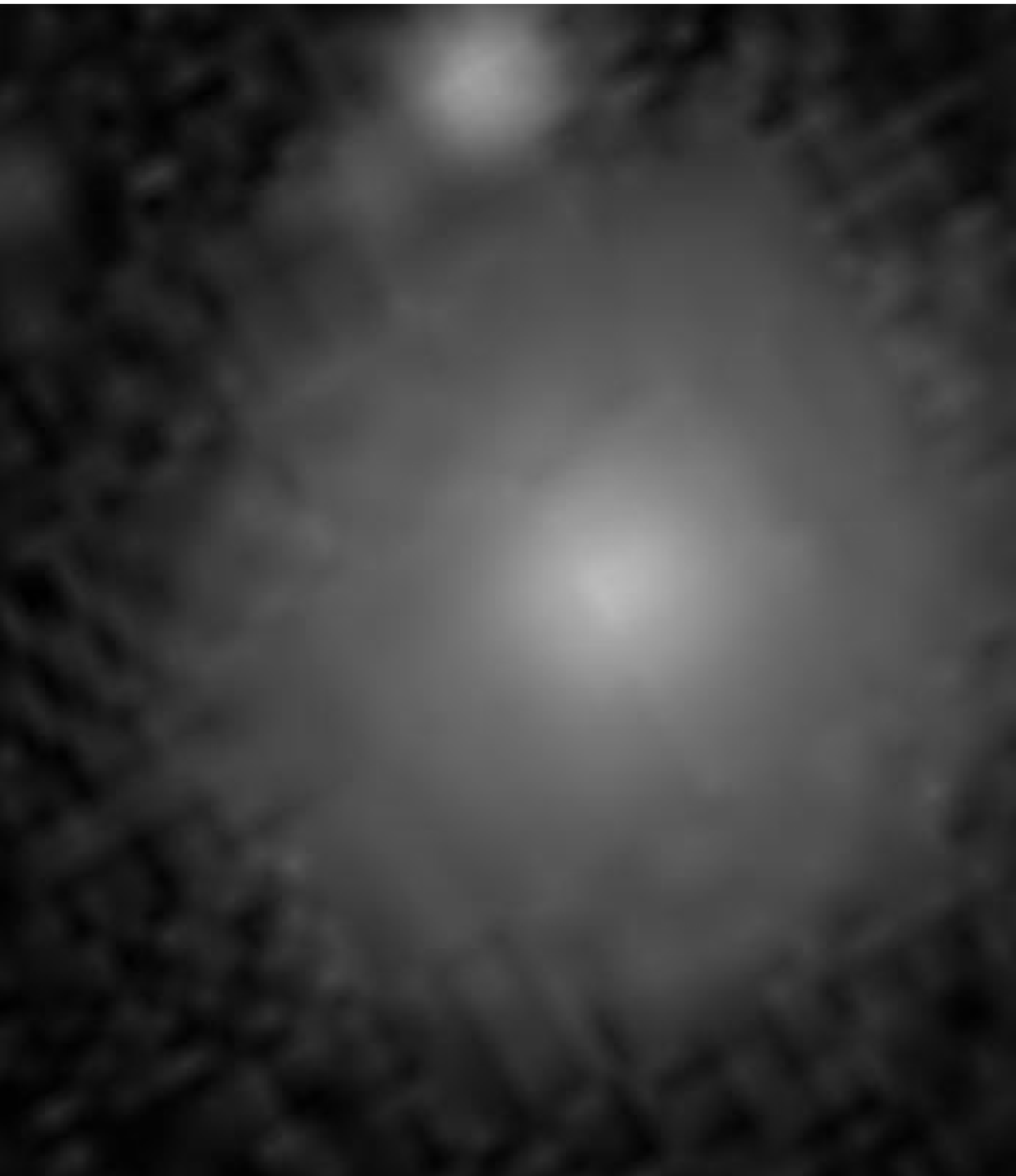}
  \caption{Spiral galaxy biased classification. Left: A spiral galaxy with good resolution taken from low earth orbit. Notice the spiral arms. Right: The same spiral galaxy, except at worse resolution and taken from the ground through the Earth's atmosphere. Notice that the arms are no longer discernible. This spiral galaxy was classified by $> 95\%$ of annotators as being an elliptical, even though higher quality data prove that it is a spiral.}
  \label{fig:S_gals}
\end{figure}

In Figure \ref{fig:S_gals} right, we show a spiral galaxy that was
classified as an elliptical with high confidence in the Galaxy Zoo
sample which is based on ground-based imaging from the Sloan Digital
Sky Survey DR7 \citep{SDSSDR7}. On the left, we show a higher
resolution view of this same galaxy from the Hubble Space Telescope,
in which one can clearly identify spiral arms. In this example, the
spiral arms are washed out by the convolution of the ground-based PSF,
rendering their structure undetectable to the human classifiers.  It
is the projected intrinsic physical scale of the underlying features
relative to the PSF that drives the misclassifications. \def\colsep{2}

In order to measure the amount of bias in different labeled data sets, we follow \cite{Bamford2009} and use the fractions of objects of each class as a function of the \emph{observable parameters} that may bias our labels. An example of such parameters for galaxy morphologies is the resolution of the galaxies: high resolution galaxies are hardly going to be mislabeled, while low resolution galaxies are more likely to be so. We would expect the fractions of an unbiased data set not to depend on these observable parameters. At the same time we should also consider \emph{intrinsic parameters} for which the real fractions of labels will depend on. 

Consider a set of {\it intrinsic} properties (e.g., physical size, luminosity, or redshift) $\boldsymbol{\beta} = \{\beta_1, \cdots, \beta_{n_\beta} \}$ on which we define $N_\mathcal{B}$ multi-dimensional bins $\mathcal{B}_q$. Given a set of $K$ labels (e.g. $K=2$ for spirals and ellipticals), in each bin $\mathcal{B}_q$, we calculate the {\it intrinsic class fraction} of objects with each label as $f_{k, q}$. For typical galaxy morphology data sets, we define $\boldsymbol{\beta}_i = (R_i, M_i, z_i)$, where $R_i$ is the physical radius (in kpc), $M_i$ is the absolute magnitude, and $z_i$ is the redshift for object $i$. In other words, given a fixed bin $q$ in galaxy physical size, luminosity, and redshift, $f_{k={\rm spiral}, q}$ defines the {\it intrinsic fraction} of spirals compared to the total number of galaxies in bin $q$.

We then consider the set of {\it observed} properties of the objects (e.g., angular size). We define the set of properties $\boldsymbol{\alpha} = \{\alpha_j\}_{j=1}^{n_\alpha}$ and create single dimensional bins on each observed property for each of the $\mathcal{B}_q$ multi-dimensional bin $\mathcal{A}_{j,l,q}$. Here $j$ defines which property and $l$ defines the range of the bin for that property. For typical galaxy morphological data sets, we define $\boldsymbol{\alpha}_i = (r_i/\mathrm{PSF}_i)$  where $r_i$ is the angular size and $\mathrm{PSF}_i$ is the estimated size of the point spread function at the galaxy location in the same units as its angular size.

\begin{figure*}
\includegraphics[width = 7 in]{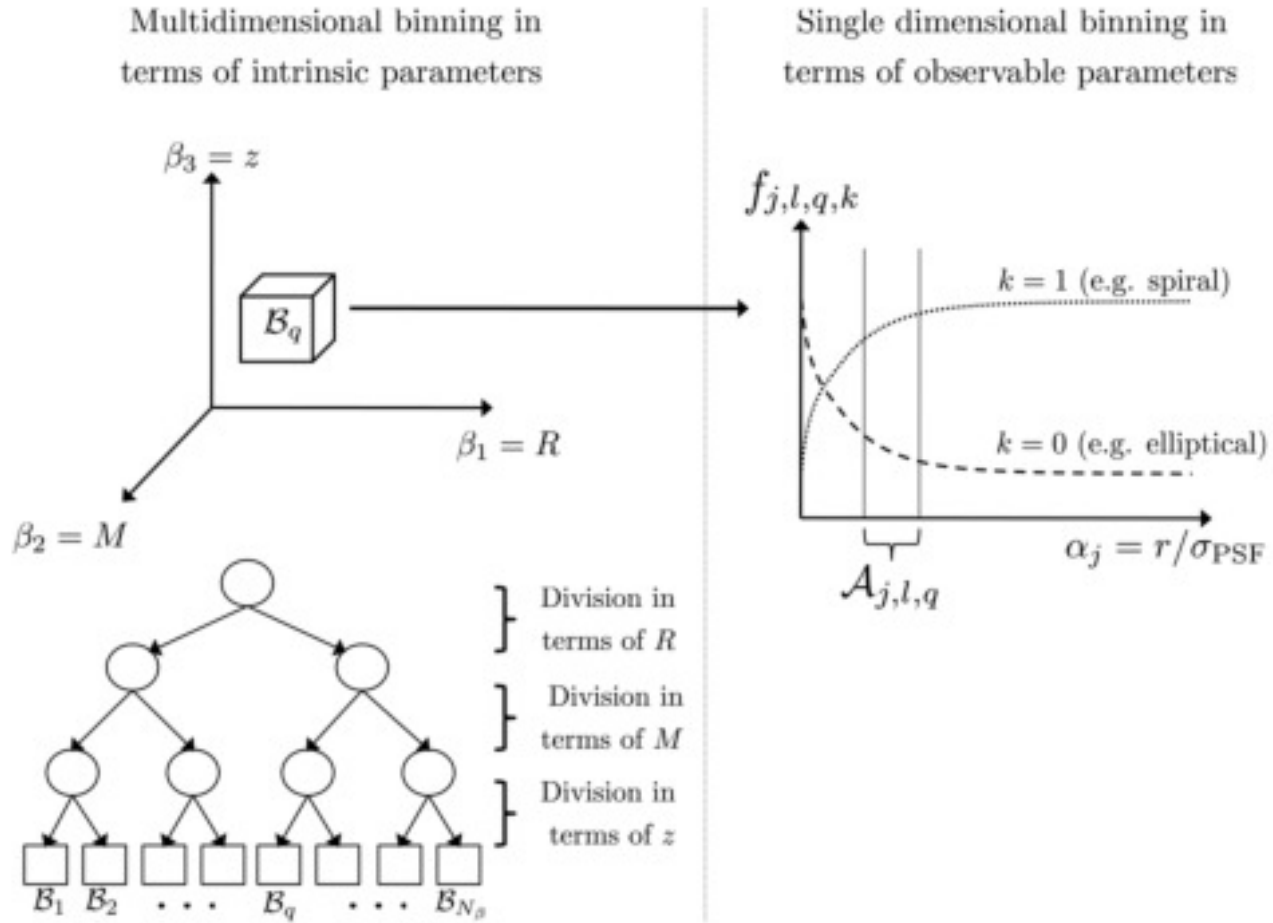}
\caption{Binning on the intrinsic and observed object properties. {\bf Top Left:} We first create multidimensional bins, $\mathcal{B}_q$, based on the intrinsic properties, such as absolute magnitude (M), physical size (R), and redshift (z). {\bf Bottom left:} These bin edges are defined using a kd-tree, as explained in Section \ref{sec:MeasuringGalaxyBias}. {\bf Right:} For the sub-set of galaxies within each intrinsic bin $\mathcal{B}_q$, we measure the fraction of labeled objects $f_{j, l, q, k}$ for each class $k$ as a function of each observable $\alpha_j$. We define equal-sized one dimensional bins on the observed properties $\mathcal{A}_{j, l, q}$, where $l$ runs over these bins and we require at least two bins. For an un-biased data set, the fractions within each bin $\mathcal{B}_q$ should not change in terms of the observables. When labeling bias is present, the fractions of objects labeled by humans will depend on the observable parameters. We calculate the deviation of fractions from  the intrinsic class fraction (Eq. \ref{eq:sigma_frac}) and then define the total labeling bias by summing over all of the properties as in Eq. \ref{eq:L}.}
\label{fig:calculatingL}
\end{figure*}

Note the intrinsic properties are treated in multi-dimensional bins, $\mathcal{B}_q$, whereas within each of those bins, the observed properties are treated in individual bins, $\mathcal{A}_{j,l,q}$. This is because our aim is to study the biases with respect to their observed individual properties and so we require at least two bins ($l\in\{1,2\}$) for each observational property. Figure \ref{fig:calculatingL} shows a diagram explaining binning in the intrisic and observable parameters. We start by defining bins $\mathcal{B}_q$ in the intrinsic parameters using a kd-tree (see Section \ref{sec:MeasuringGalaxyBias}). For each of this multi-dimensional bins we bin again in terms of the observable parameters and calculate the fraction of objects in each of these bins for every class. 

We then calculate the {\it observed class fraction}
\begin{eqnarray}
f_{j, l, q, k} = 
\frac{1}{N_{\mathcal{A}_{j,l,q}}}\sum_{\substack{
i|\alpha_{i,j}\in \mathcal{A}_{j,l,q}\\
\boldsymbol{\beta}_i\in \mathcal{B}_q}}\delta_{\hat{y}_i,k}, \label{eq:frac_klq}
\end{eqnarray}
where  $N_{\mathcal{A}_{j,l,q}}$ is the total number of objects with the observed property $\alpha_j$ in bin $\mathcal{A}_{j,l,q}$,
$\delta_{\hat{y}_i,k}$ is the Kronecker delta given an estimate of each galaxy $i$'s classification $\hat{y}_i$ for class $k$. The right-hand sides sums over all galaxies which are simultaneously in the observed single property bin $\mathcal{A}_{j,l,q}$ and the intrinsic property multi-dimensional bin $\mathcal{B}_q$. 

For a given classification $k$ and intrinsic property bin $\mathcal{B}_{q}$, we calculate the $\ell^2-$ Euclidean difference between the observed class fraction $f_{j, l, q, k}$ and the intrinsic class fraction $f_{k, q}$ and sum over all the $N_{\mathcal{A}_{j,q}}$ bins $\mathcal{A}_{j,l,q}$ for the observed property $\alpha_j$
\begin {equation}
\sigma_{j,k,q}^2 = {\frac{1}{N_{\mathcal{A}_{j,q}}}\sum_{l=1}^{N_{\mathcal{A}_{j,q}}}(f_{j, l, q, k} - f_{k, q})^2}
\label{eq:sigma_frac}.
\end{equation}
Equation \ref{eq:sigma_frac} should be $\sim 0$ for large $N$ and when there is no difference between the intrinsic and observed class fractions, i.e., when the classifications are unbiased with respect to an observable.

We can extend this to all classes and intrinsic and observed properties as
\begin {equation}
L = \sqrt{\frac{1}{KN_\mathcal{B}n_\alpha}\sum_{j,k,q}{\sigma_{j,k,q}}^2},
\label{eq:L}
\end{equation}
where $K$ is the number of classes (2 for the case of elliptical versus spirals). 
We term equation \ref{eq:L} the classification bias which quantifies the difference in the {\it observational class fractions} with respect to the {\it intrinsic class fractions}. 

We note that the {\it intrinsic class fraction} $f_{k,q}$ can vary for any data set. For instance, a data set designed to represent ellipticals might have an inherently lower spiral fraction than a broader morphological catalog containing spirals, ellipticals and irregulars. Alternatively, one might be interested in comparing classification algorithms over a wide range of classes and data sets. If so, care has to be taken so that the intrinsic parameters distributions are similar so we do not have any
selection effects which could influence $f_{k,q}$ and in turn the value of $L$. 

One would hope that the fraction of labels within bins of intrinsic properties, $f_{k, q}$ could in principle be measured using an un-biased (``gold standard'') data set or perhaps a subset of the data itself. It is also possible that $f_{k,q}$ could be predicted from theory \citep[e.g. ][]{Genel2014}. Here, we take a conservative approach and assume that {\it all} observed morphological data sets have some level of bias. We make an estimate $\hat{f}_{k,q}$ by using the observed class fraction $f_{j,l,k,q}$ for the bin $l$ in observed property $j$ which is likely to have the least bias. For example, if we are calculating $\sigma_{j,k,q}$ for $\alpha_j = r/\sigma_{\rm{PSF}}$, then we calculate $\hat{f}_{k,q}$ for the bin which includes the largest values of $ r/\sigma_{\rm{PSF}}$, since it should contain the least biased classifications. 

\begin{figure*}
\includegraphics[width = 7 in]{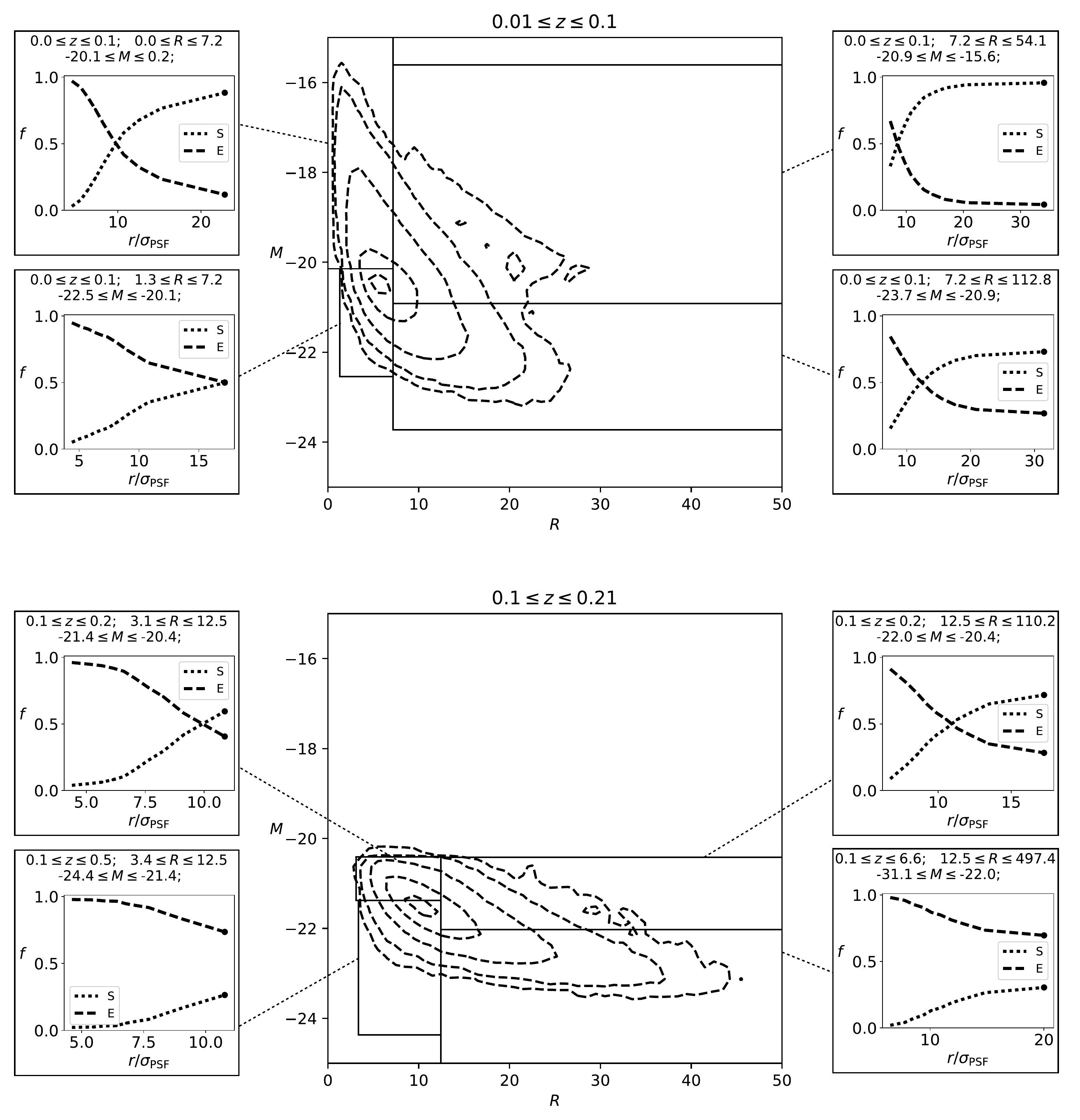}
\caption{Binning example for Galaxy Zoo biased data. The kd-tree splits the data in terms of the intrinsic properties $z$ (center top and bottom), $R$, and $M$ (solid line rectangles). For each of these 3-dimensional bins we calculate the fractions of objects in terms of the observable parameter $r/\sigma_\mathrm{PSF}$. As the size of the galaxies diminishes the fraction of spiral galaxies (dotted lines) decreases, and the fraction of ellipticals (dashed lines) increases. The least biased bins in observable parameters are the ones with the highest $r/\sigma_\mathrm{PSF}$, represented by a dot in the plots. We use these lowest bias bins as our estimation for the intrinsic fractions $\hat{f}_{k, q}$.}
\label{fig:fractions}
\end{figure*}

\begin{figure*}
\includegraphics[width = 7.2 in]{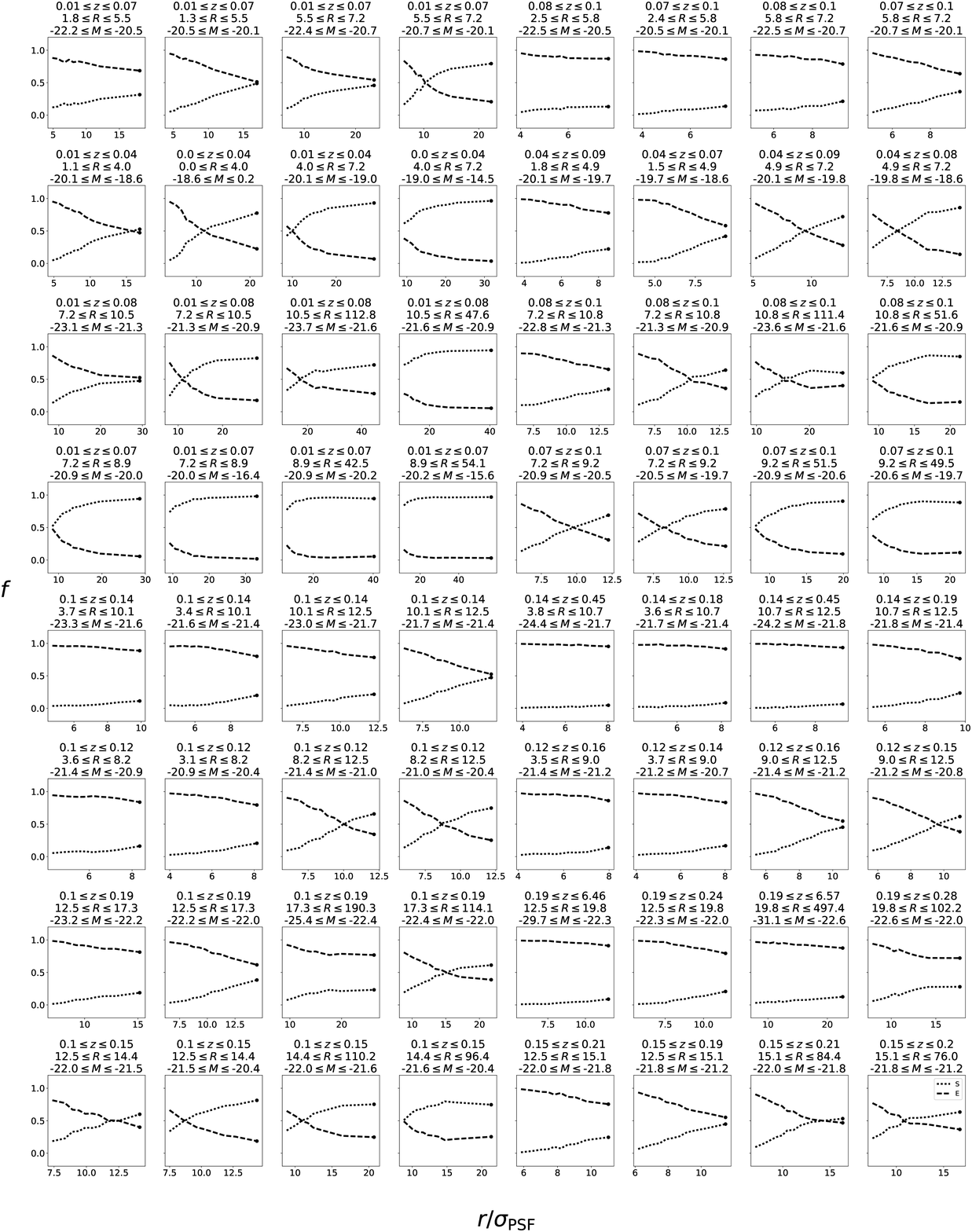}
\caption{Fractions in terms of observable parameters for Galaxy Zoo biased data using $2^6$ bins in intrinsic parameters. As the angular size of the galaxies diminishes the fraction of observed spiral galaxies (dotted lines) decreases, and the fraction of ellipticals (dashed lines) increases due to observational bias. The least biased bins in observable parameters are the ones with the highest $r/\sigma_\mathrm{PSF}$, represented by a dot in the plots, which we use as our estimate for the intrinsic fractions $\hat{f}_{k, q}$.}
\label{fig:fractions_64}
\end{figure*}

Figure \ref{fig:fractions} shows an example of binning in intrinsic and observable parameters for Galaxy Zoo data. Here, we build the kd-tree splitting the data in terms of $z$, $R$, and $M$ creating a 3-dimensional partition of the data. For the data falling in each of the intrinsic bins we calculate the fraction of spiral and elliptical galaxies as a function of the observable parameter $r/\sigma_\mathrm{PSF}$. As  $r/\sigma_\mathrm{PSF}$ decreases (smaller objects), the fraction of spiral galaxies decreases and the fraction of elliptical galaxies increases. In other words, smaller spiral galaxies are confused as elliptical. In order to calculate our bias metric (Eq. \ref{eq:L}) we need the intrinsic class fractions $f_{k,q}$. The least biased bins in the observable parameters are the ones with the biggest $r/\sigma_\mathrm{PSF}$, which we consider as our estimate for the intrinsic class fractions. Figure \ref{fig:fractions_64} shows the fractions of spiral and elliptical galaxies in term of the observable parameter $r/\sigma_\mathrm{PSF}$ for $2^6$ bins in intrinsic parameters. Independently of the bin in terms of $z$, $R$, and $M$, the fraction of spirals increases with $r/\sigma_\mathrm{PSF}$, while the fraction of ellipticals decreases. In order to calculate the dataset bias, we use as intrinsic class fractions $f_{k,q}$ the fraction in the bin with a higher $r/\sigma_\mathrm{PSF}$, denoted by a dot in Figure \ref{fig:fractions_64}.

\section {data sets}
\label{sec:data}

In this section we describe the data we use on our experiments. All data considers the r-band from SDSS \citep{SDSSDR7}, and the 9-year WMAP cosmology \citep{Hinshaw2013} from astropy \citep{astropy2018}.

\subsection {Eyeball Classifications}
\cite{Fukugita2007} (hereafter F07) have visually classified 2,275 galaxies, each by three experts. They defined a morphological index $T$ such that $T = 0, 1, 2, 3, 4, 5, 6$ for E, S0, Sa, Sb, Sc, Sd, Im, respectively. In order to measure their bias, we focus on just the elliptical (+S0) galaxies (N=941) having $0 \leq T < 2$, and the spirals (N=902) having $2 \leq T \leq 5$, since the other data sets we compare to only use these two classes. We cross-match these data to the
SDSS DR7 to obtain their apparent magnitudes (Petrosian r-band), their
apparent sizes (Petrosian r-band radii), and their redshifts.

We also use expert labels from \cite{Nair2010} (NA10 hereafter) who have visually classified 14,034 spectroscopically-targeted galaxies from the SDSS. They report T-Types as well as other morphological features such as bars, rings, lenses tails, among others. As with the F07 sample, we focus on elliptical (+S0) galaxies (N=6,276) having $-5 \leq TT < 1$ and spirals (N=7,640) having $1 \leq TT \leq 8$, where $TT$ are their T-Types.


\subsection {Galaxy Zoo}
We use the Galaxy Zoo 1 data release \citep{Lintott2011} and their sample with spectra in SDSS which contains classifications for 667,944 galaxies achieved by crowd-sourcing. We define two subsets of the Galaxy Zoo 1: (a) the original biased morphologies (hereafter GZB) and (b) the ``debiased'' morphologies (hereafter GZD). The debiasing procedure used is described in detail in \citet{Bamford2009} and \citet{Lintott2011}. Briefly, they assumed that the morphological fraction within bins of fixed galaxy physical size and luminosity does not evolve over the redshift of their data. From that assumption, a bias correction term was estimated in bins of physical size and luminosity and then applied to the original spiral and elliptical classification probabilities. Their algorithm helped motivate our approach to quantify classification bias as described in Section \ref{sec:bias}.

We cross-match the Galaxy Zoo catalog to the SDSS DR7 to obtain the observed properties, including each galaxy's point-spread function (PSF-determined over the SDSS field).  We used the
SDSS field-specific $\rm{psfWidth\_r}$ parameter as an estimate of the
FWHM for a Gaussian PSF at the location of each galaxy. 
When galaxies belong to more than one field we used the galaxy classification and properties pertaining to that with the smallest PSF.

\subsection {Supervised Learned Morphologies}
\cite{Huertas2011} (hereafter HC11) used a {\it support vector machine (SVM)} classification model trained over the data set from \cite{Fukugita2007}. The HC11 morphologies are probability densities, and so we defined elliptical (+S0s) galaxies as having a probability of being early-type $P($Early$) \geq p$ and spiral galaxies having $P($Spiral$) \geq p$, where $p$ takes values of 0.5 and 0.8. As with the previous data sets, we cross-match the HC11 data to the SDSS DR7 to ensure that all galaxies in our data sets have the same observed properties and that there are no duplicates.

\subsection {Simulated Morphology Catalogs} \label{sec:simulations}

In order to assess the validity of our method, we created a simulated catalog following the Galaxy Zoo 1 distribution of parameters. We used a kernel density estimation \citep[see][and references therein]{Hastie2009} with a Gaussian kernel to estimate the distribution of angular Petrosian radius $r$, apparent Petrosian magnitude $m$, redshift $z$, PSF, and de-biased probabilities randomly choosing 100.000 galaxies from GZ1. Using these parameters we calculate their physical Petrosian radius $R$, absolute Petrosian magnitude $M$, and $r_{\rm{PSF}} = r/\sigma_{\rm{PSF}}$. 
We consider $r_{\rm{PSF}}$ and $m$ as the biasing parameters, so we artificially created this bias by changing the labels from spirals to ellipticals with a Gaussian probability depending on these parameters:
\begin{equation}
P(\hat{y} = \mathrm{E}| y = \mathrm{S}) = \exp{\left(-\frac{1}{\theta^2}\left[\frac{r_{\rm{PSF}}^2}{2\bar{r}_{\rm{PSF}}^2} + \frac{(m-17.8)^2}{2(\bar{m}-17.8)^2}\right]\right)},
\end{equation}
where the probability of modifying a label from S to E depends of a biasing parameter $\theta$ which controls the amount of bias in the data set. The higher the value of $\theta$, the larger the amount of bias. Notice that this added bias is normalized in terms of $r_{\rm{PSF}}$ and $m$, by using their median values $\bar{r}_{\rm{PSF}}$, and $\bar{m}$.

\section{Impact of Sampling over the Estimator}\label{sec:MeasuringGalaxyBias}

Equation \ref{eq:L} is a statistical measure of the classification bias for any data set with $K$ classes and requires bin definitions on the observed properties $\alpha_j$ and multi-dimensional data set binning of the intrinsic properties $\beta_j$. In this section, we examine the effects of how the bins are defined using the simulated morphology catalogs which have varying degrees of bias. 

We bin the intrinsic properties of the data using kd-trees. A kd-tree is a data structure for storing a finite set of points from a k-dimensional space. It was
examined in detail by \cite{Bentley1975} and \cite{Friedman1977}. kd-trees have the benefit of  dividing the data into bins for optimal querying performance. They are well characterized in the literature and numerous libraries exist to build such trees. The total number of bins in these trees is $2^n$, where $n$ is the height of the tree. As $n$ increases, the bins get smaller causing the number of points inside each bin to get smaller too. The dimension of our kd-tree depends on the number of intrinsic properties $n_\beta$ we are examining. In this effort, we use the absolute magnitude, the physical size, and the redshift for our tree.  

For the observed properties, we need to build a grid defining the ranges on each of the $\alpha_j$ observational parameters (e.g., such as the resolution limits within the bin). We choose a simple linear binning procedure such that the number of observed galaxies in each bin is equal.

Having defined the bins on the intrinsic and observed properties of each galaxy, as well as the morphological classifications, we examine the robustness of the labeling bias estimator, Equation \ref{eq:L}.

\subsection {Finite-Sampling Bias and Variance}
\label{sec:sample_bias}

The number of 3D bins on the intrinsic properties, the number of 1D bins in the observable parameters, as well as the total number of objects in each of the bins, combine to impact $L$. Because real data sets have finite size, the trade-off between bias and variance of our estimators has to be taken into account when defining the binning strategy. We use the simulations to show the impact of the selected binning strategy on our bias metric.
Our results are shown in Figure \ref{fig:Variance}  where the left panel is for simulated galaxies following GZD probability distributions ($\theta = 0.0$) and the right panel shows a simulated bias of $\theta = 1.0$.

\def\imWidth{3.3}
\begin{figure*}
\centering
\begin{tabular}{cc}
un-biased simulations & bias $\theta = 1.0$\\
\includegraphics[width = \imWidth in]{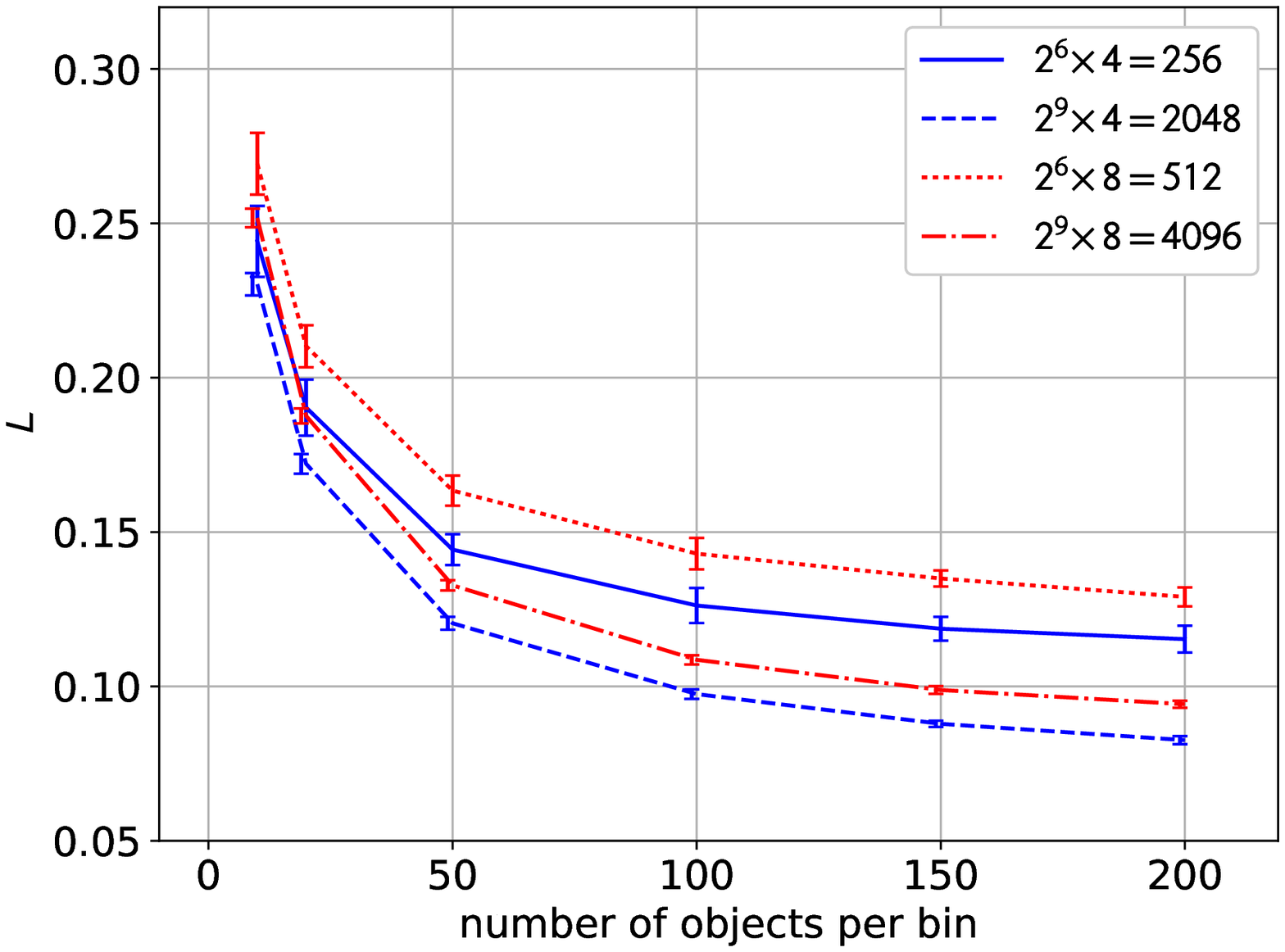} & \includegraphics[width = \imWidth in]{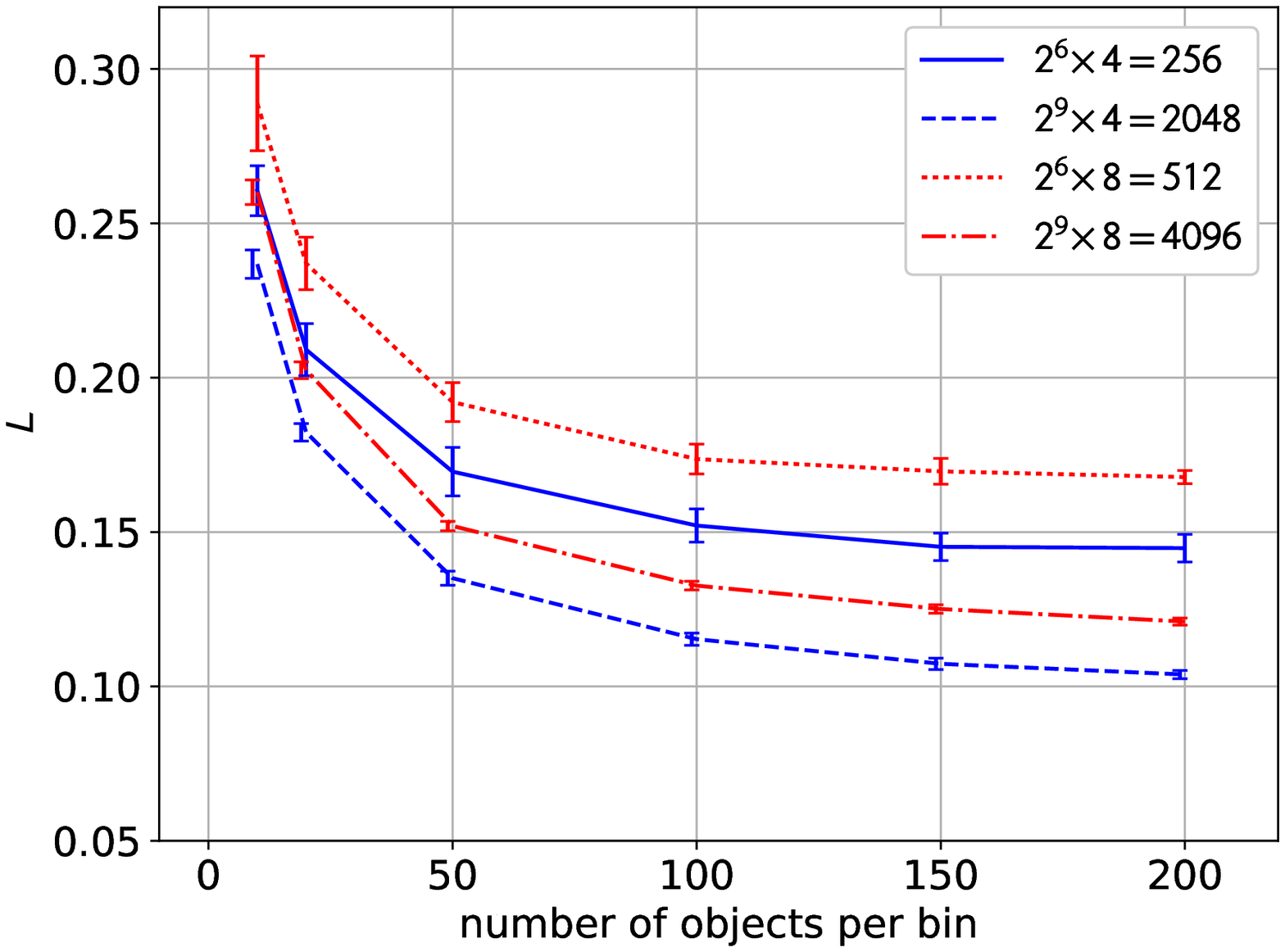}
\end{tabular}
\caption{Sampling effect over $L$ as calculated over the simulated data sets. Left: un-biased simulations. Right: bias $\theta = 1.0$. Labels indicate the number of bins in the intrinsic parameters obtained from the KD-tree, and the number of bins in the observable parameter: $N_\mathcal{B}\times N_{\mathcal{A}_{j, q}}$. The variance over $f_{j,l,k,q}$ increases as we diminish the number of objects per bin increasing the value of $L$. Independently of the binning strategy chosen, our metric obtains a higher value for the biased simulated dataset.}
\label{fig:Variance}
\end{figure*}

First, consider a simple fixed binning scenario where we allow the number of galaxies per bin to vary. 
Figure \ref{fig:Variance} shows this effect over simulations for different binning strategies. One can see that $L$ decreases as a function of the square root of the total number of galaxies in each bin. There is a point after which adding more galaxies to each bin does not reduce $L$ significantly. We use the shape of this curve to define the optimal number of total galaxies per bin.

Next, consider a fixed number of objects per bin and a fixed number of bins on the intrinsic properties. 
As one decreases the number of 1D bins in the observable parameters, due to the bias-variance trade-off \citep[see][]{Hastie2009}, there will be a corresponding decrease in the statistical variance for the estimate of the fractions $f_{j,l,k,q}$, at the expense of increasing the statistical bias. The extreme case is a single bin with very low variance. However, as shown in Figure \ref{fig:calculatingL}, a single bin in $\mathcal{A}_{j, l, q}$ provides no useful information on the bias we are trying to measure: at least two bins in the observed properties are required in order to track observational bias. Regardless, the decrease in variance (simply due to fewer bins) simultaneously decreases the value of the labeling bias $L$. This is shown in Figure \ref{fig:Variance} by considering curves with the same number of intrinsic bins and noting that $L$ is always lowest for the fewest number of observed bins.

On the other hand, if we fix the number of objects per bin as well as the number of bins on the observed properties, then by decreasing the number of bins on the intrinsic parameters, we lose information about the true object fractions, thus causing an increase in the bias of the estimator for the intrinsic class fraction $\hat{f}_{k,q}$. This produces an increase of the differences between the observational class fractions and the estimate for the intrinsic class fractions in Eq. \ref{eq:sigma_frac} increasing the value of $L$.
This is shown in Figure \ref{fig:Variance} by considering curves with equal number of bins in the observables noting that $L$ is always lowest for the highest number of intrinsic bins.

Figure \ref{fig:Variance} allow us to define a binning procedure for any data set. Notice how the number of objects per bin and binning impacts the value of our bias metric $L$ for data sets with the same amount of simulated bias. Also notice that $L$ is always higher for the data set with higher simulated bias $\theta$ for a given binning strategy, which suggests that any binning strategy helps evaluate differences in biases between data sets, as long as enough number of objects per bin are considered.

Since the value of $L$ can vary as a function of the binning, we must be careful to use the same binning procedure when conducting relative comparisons of one or more data sets, even if the binning is not optimal for any specific data set. In practice, when comparing the labeling bias for different sets of data, the binning strategy is defined by the data set with the smallest number of objects.

\subsection{Choosing number of bins for real data}\label{sec:choosing_bins}

For the datasets in this work, we consider binning strategies that split all parameters (observable and intrinsic) into the the closest number of bins. Given this constraint, we then search for the maximum number of bins such that the running slope of $L$ in Figure \ref{fig:Variance} is $< 10^{-3}$ for the maximum number of objects per bin allowed by our data set size.
Because the real data is noisy, we calculate the mean value of $L$ over 20 bootstrapping sub-samples and considering the same number of bins for each intrinsic and observable parameters. The kd-tree automatically defines the multidimensional binning on the intrinsic parameters. 

Special care has to be taken when comparing two or more data sets of different sizes. On a larger data set we may be able to use more bins and/or number of objects per bin, but when comparing it to a smaller data set, this sampling is not going to be feasible to use. In order to make a fair comparison, we need to sample in terms of the \emph{smaller data set}, so that biases and variance over the distribution of fractions are comparable.


\section {Bias for galaxy morphologies}\label{sec:GalaxyBias}

Now that we defined how to choose the binning in Equation \ref{eq:L}, we measure the classification bias $L$ for the different datasets defined in Section \ref{sec:data}. In Section \ref{sec:z_obs} we follow the approach proposed by \cite{Bamford2009} and use the redshift as a way  to quantify the morphological bias. Then, in Section \ref{sec:rpsf_obs} we use $r/\sigma_{\mathrm{PSF}}$ as our biasing observable parameters and $R$, $M$, and $z$ as intrinsic parameters.

\subsection{Redshift as biasing parameter} \label{sec:z_obs}
\begin{figure*}
\centering
\begin{tabular}{ccc}
\includegraphics[width =0.33 \textwidth] {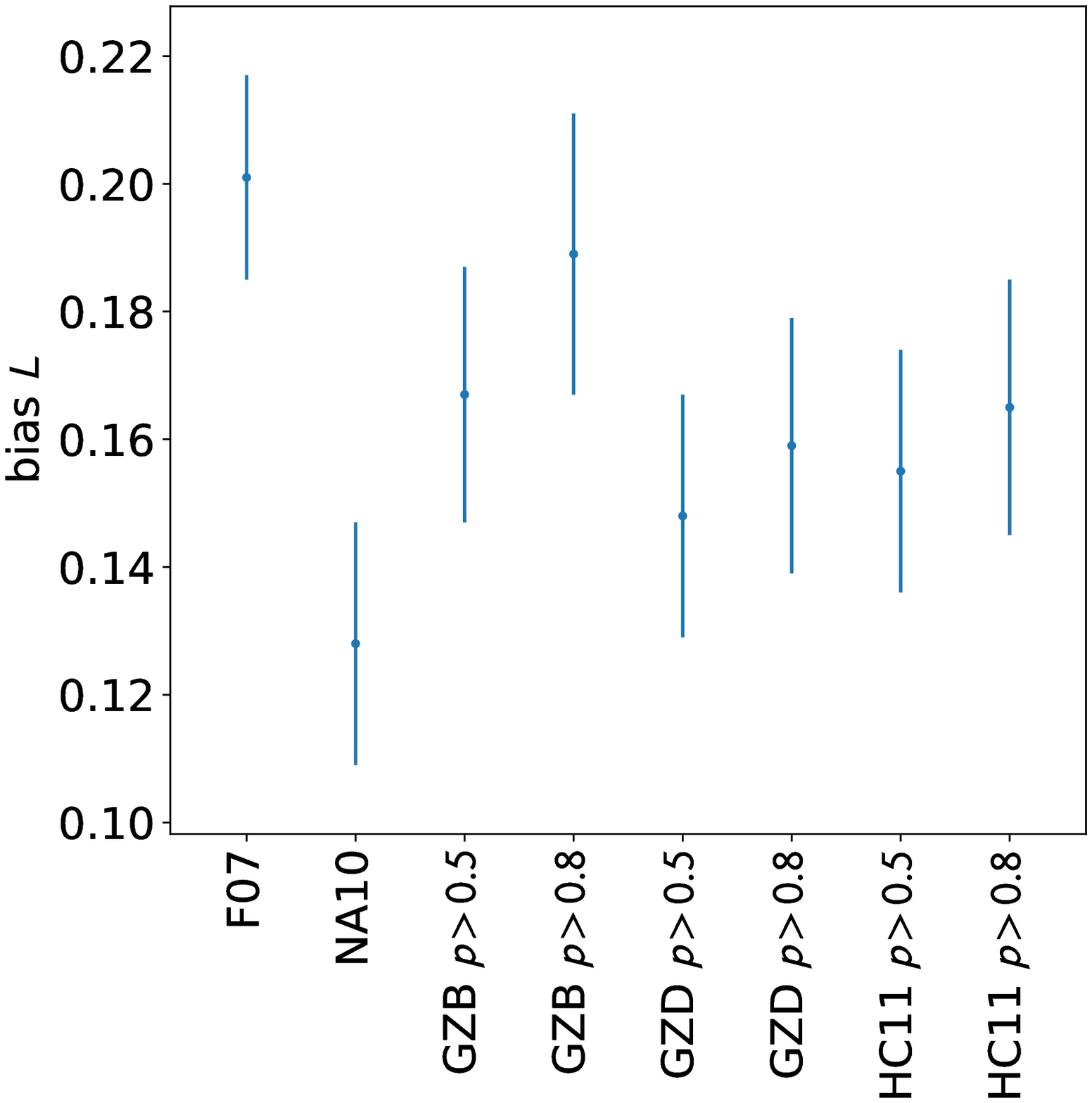} &
\includegraphics[width =0.33 \textwidth] {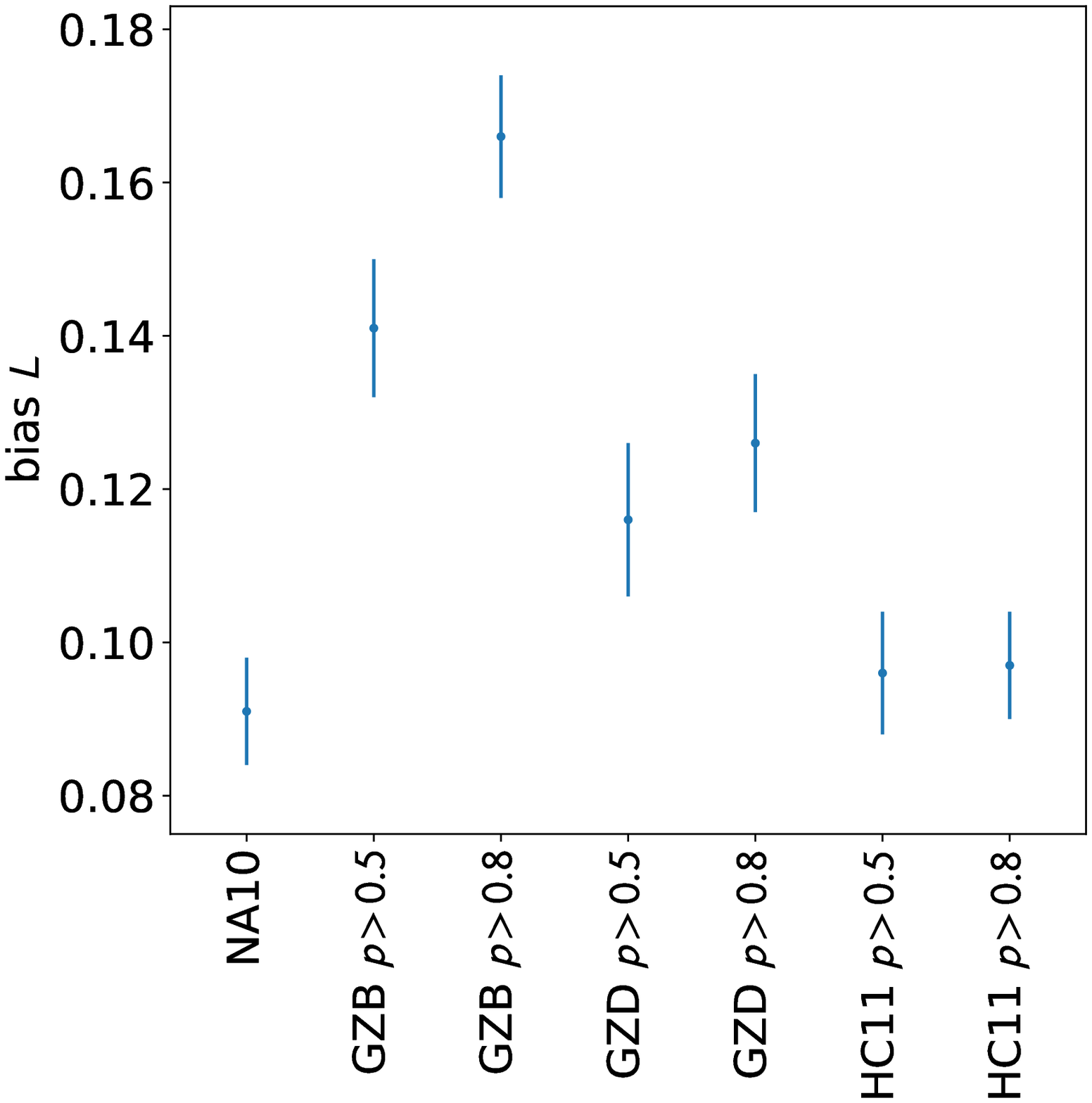} &
\includegraphics[width =0.33 \textwidth] {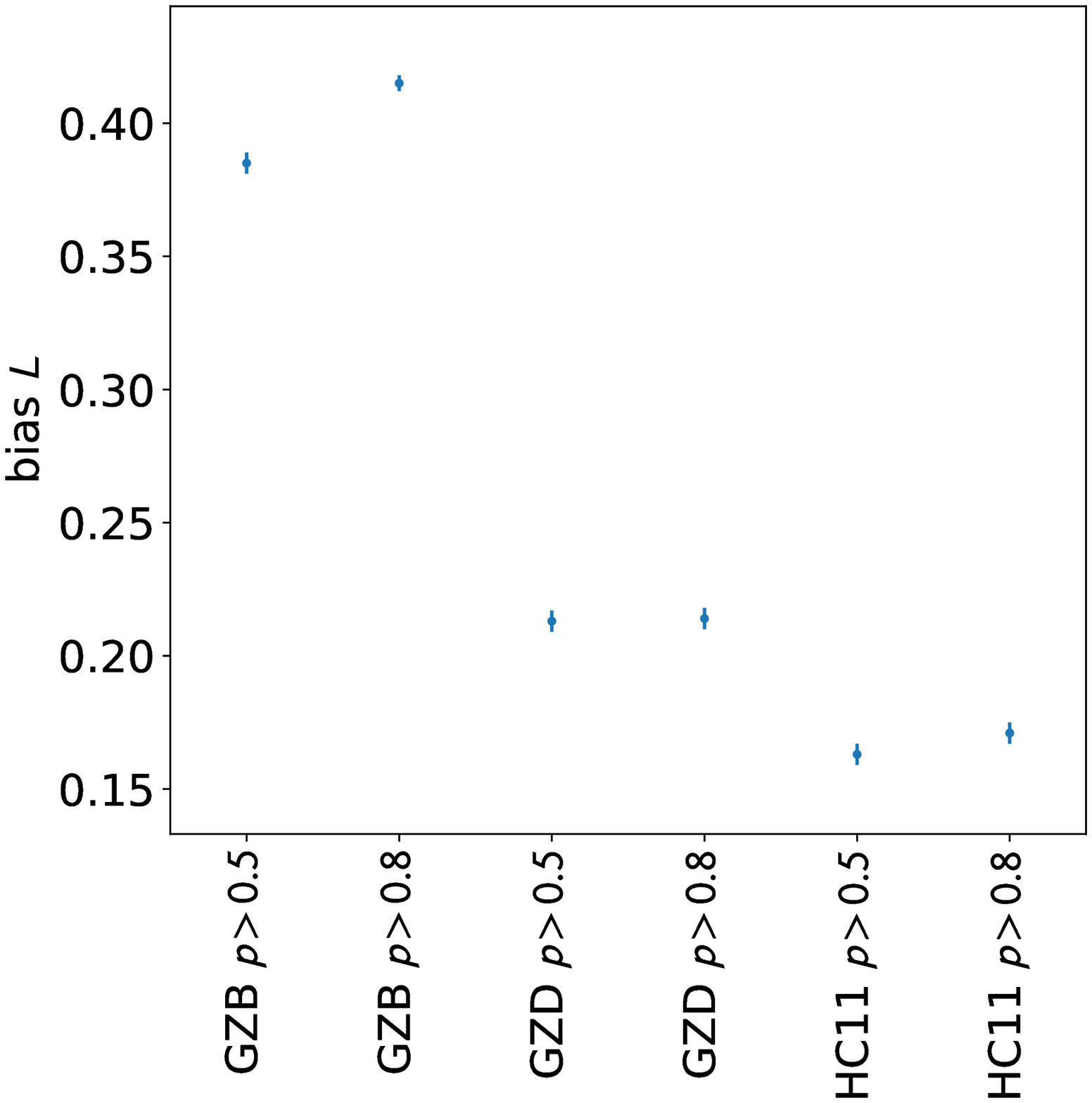} \\
(a) & (b) & (c)\\
\end{tabular}
\caption{Bias $L$ for different datasets considering $z$ as unique observable parameter and $R$ and $M$ as intrinsic parameters, as proposed by \cite{Bamford2009}. The number of galaxies considered for measuring the bias increases from (a) to (c). Error bars show the standard deviation over 100 bootstrapping samples. (a) Using $2^3$ bins in intrinsic parameters, $2$ bins in observable parameters, and 115 galaxies per bin in order to match the total number of galaxies of \cite{Fukugita2007} (F07). (b) Using $2^4$ bins in intrinsic parameters, $4$ bins in observable parameters, and 217 galaxies per bin in order to match the total number of galaxies of \cite{Nair2010} (NA10). (c) Using $2^8$ bins in intrinsic parameters, $16$ bins in observable parameters, and 58 galaxies per bin in order to match the total number of galaxies in the Galazy Zoo Biased (GZB) sample \cite{Bamford2009}. Note that the machine learning classifications of \cite{Huertas2011} (HC11) use the F07 classifications for training the Support Vector Machine.}
\label{fig:bias_Bamford}
\end{figure*}

We start by following the approach proposed by \cite{Bamford2009} and consider redshift as a biasing parameter and physical radius and absolute magnitude as intrinsic parameters. The smallest dataset is F07 with 1843 spirals and ellipticals. From this, we used the technique described in Section \ref{sec:choosing_bins} to determine the binning. We find the best finite-sampling bias levels for a maximum binning size of 8 in the intrinsic parameters and 2 in the observed parameters. We then apply this binning scheme to all of the datasets to measure the classification bias using Equation \ref{eq:L}. For the F07 data set we obtain 115 galaxies per bin, so we fix this number for all the other datasets. The data from F07 and NA10 only contains galaxies for $m < 16$, so in order for the comparison of biases between datasets to be fair, we consider galaxies with $m < 16$ in the GZ and HC11 datasets. Figure \ref{fig:bias_Bamford}a shows the bias for different datasets. Notice the standard deviation of $L$ makes it hard to make statistical significant conclusions on the difference between datasets.
 
If we exclude F07, the smallest dataset is NA10, with 13,916 galaxies. For these data, and using the procedure from Section \ref{sec:choosing_bins}, we find a maximum binning size of $2^4$ in the intrinsic parameters and 4 bins in the observable parameters (217 galaxies per bin). Figure \ref{fig:bias_Bamford}b shows the results of the bias for each data set. The error bars now allow us to interpret these results with higher significance. The bias for expert annotators of  NA10 is similar to the one from HC11, and both are smaller than the bias of Galaxy Zoo. Now, we can see that our metric starts to recover the de-biasing procedure from \cite{Bamford2009}: the value of $L$ is lower for GZD than for GZB.

If we exclude both F07 and NA10, we can consider galaxies with $m>16$. The smallest dataset is the Galaxy Zoo biased with 237,963 galaxies, so by doing this, we are able to use a larger number of bins, thus having best estimates for $L$. Using the method described in Section \ref{sec:choosing_bins} we obtain $2^8$ bins for the intrinsic parameters and 16 bins for the observable parameters, from which we can use 58 objects per bin. In Figure \ref{fig:bias_Bamford}c we show the labeling bias as defined by Equation \ref{eq:L} using this binning strategy for GZB, GZD, and HC11. Now, we can clearly recover the de-biasing procedure proposed by \cite{Bamford2009}. The highest values of $L$ are obtained over the GZB dataset, while the GZD dataset achieves a significantly lower $L$. This shows, that our proposed metric is capable of measuring biases given an assumption of intrinsic and biasing parameters. Again, the lowest labeling bias is obtained for HC11.

\subsection{Apparent radius as biasing parameter} \label{sec:rpsf_obs}
As opposed to Bamford et al. who utilized redshift as the parameter for which to characterize and correct labeling bias, in this section we treat the apparent size as the parameter which governs bias. With respect to the PSF, it is the apparent size of a galaxy that will determine whether or not spiral features are washed out to become undetectable. We then include redshift as an intrinsic parameter since we expect it to play a role in the underlying fraction of spirals and ellipticals, which we know to evolve over time \citep{Buitrago13, Huertas2015b, cerulo2017}. There is a concern that the apparent size as an observable parameter is degenerate with the combination of the redshift and the physical size for any galaxy. A small nearby galaxy can have the same apparent size as a large and more distant galaxy. However, by also including the absolute magnitude as an intrinsic parameter, this degeneracy is broken. In other words, a small and large galaxy with the same apparent size will never be in the same bin since the small (and thus intrinsically dim) galaxy will appear in a different Magnitude bin than a large (and intrinsically bright) galaxy.

\begin{figure*}
\centering
\begin{tabular}{ccc}
\includegraphics[width =0.33 \textwidth] {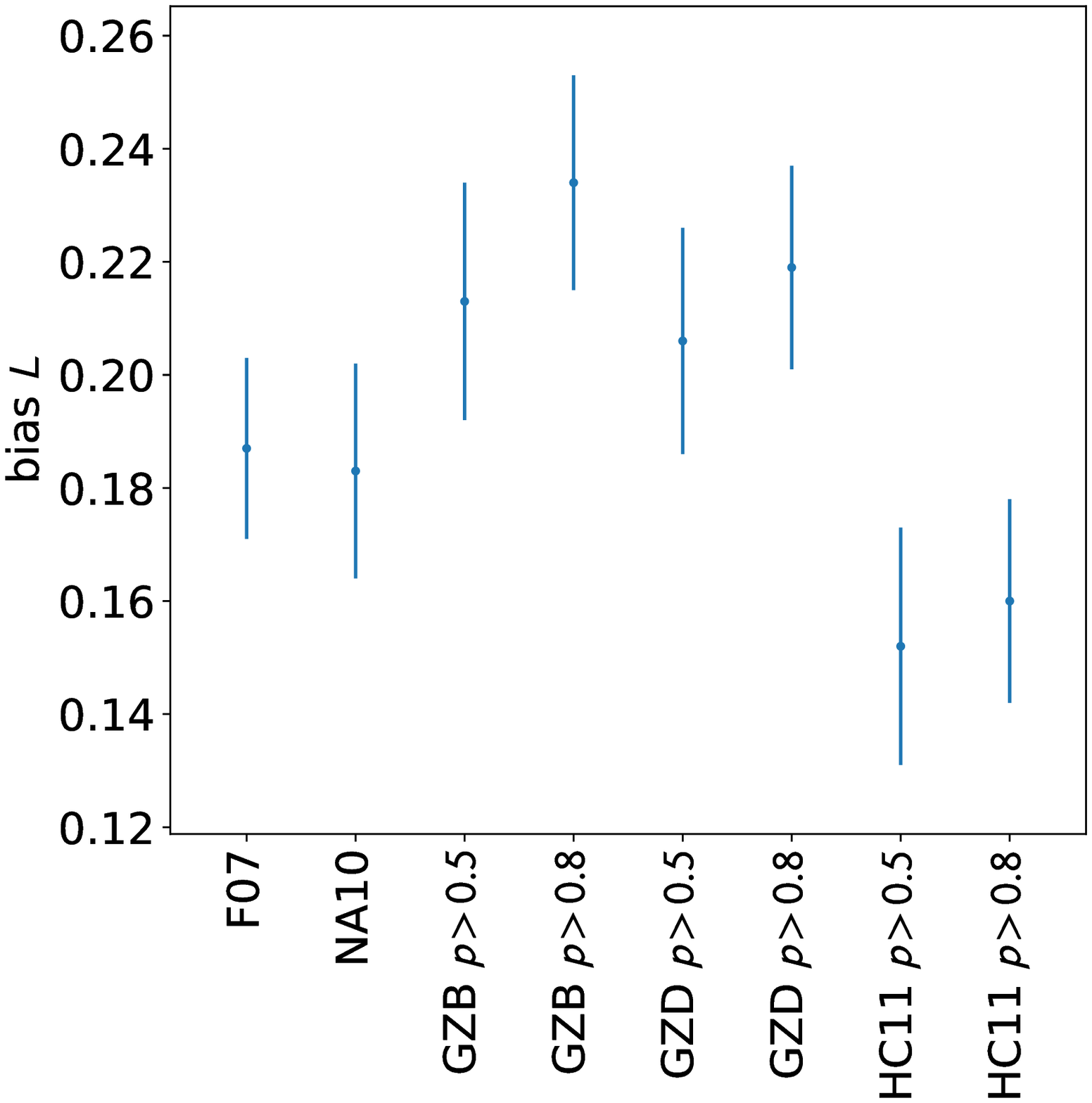} &
\includegraphics[width =0.33 \textwidth] {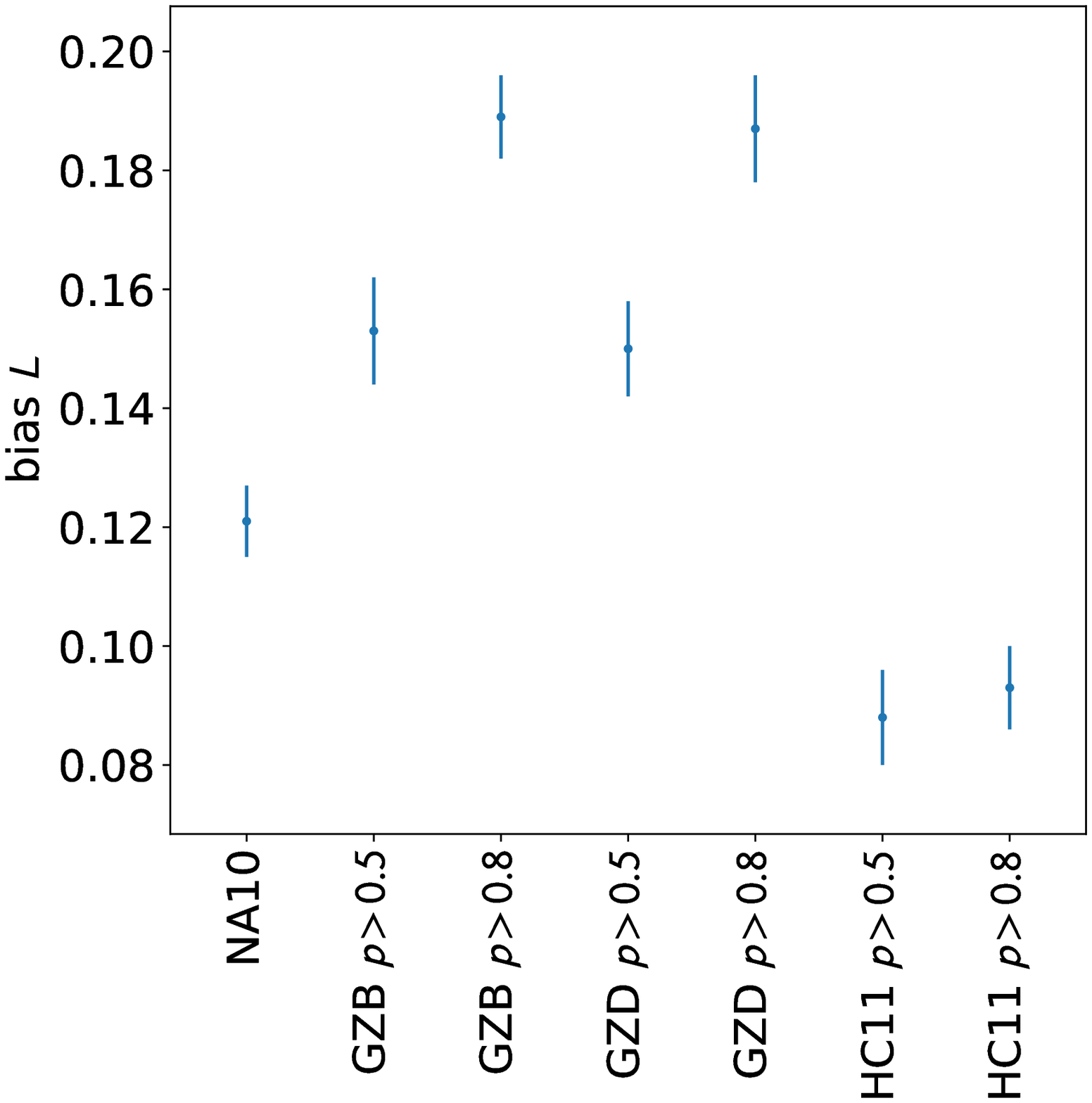} &
\includegraphics[width =0.33 \textwidth] {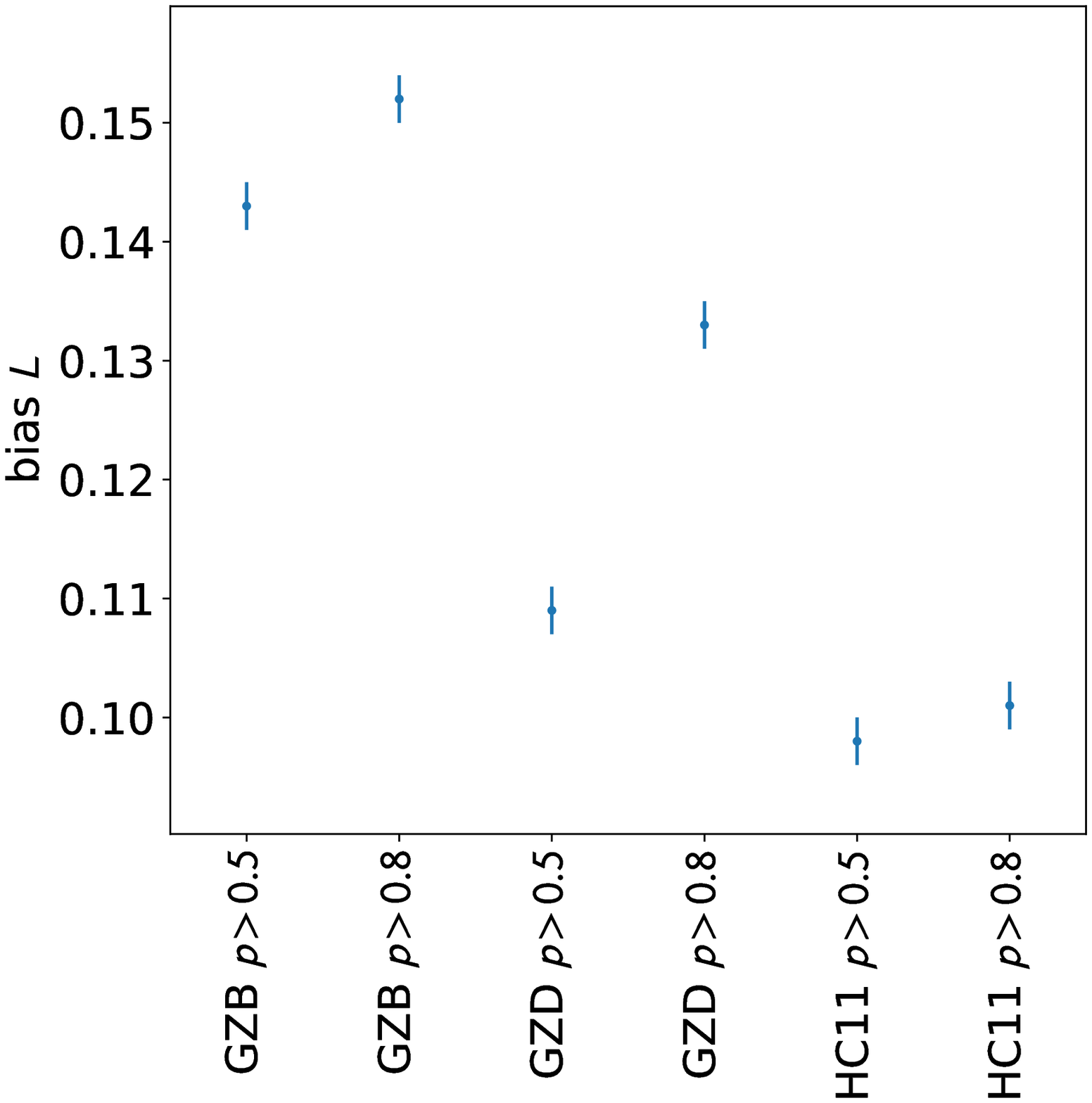} \\
(a) & (b) & (c)\\
\end{tabular}
\caption{Bias $L$ for different datasets considering $r_\mathrm{PSF}$ as unique observable parameter and $R$, $M$, and $z$ as intrinsic parameters. The number of galaxies considered for measuring the bias increases from (a) to (c).  Error bars show the standard deviation over 100 bootstrapping samples. (a) Using $2^3$ bins in intrinsic parameters, $2$ bins in observable parameters, and 115 galaxies per bin in order to match the total number of galaxies of \cite{Fukugita2007} (F07). (b) Using $2^5$ bins in intrinsic parameters, $3$ bins in observable parameters, and 144 galaxies per bin in order to match the total number of galaxies of \cite{Nair2010} (NA10). (c) Using $2^8$ bins in intrinsic parameters, $4$ bins in observable parameters, and 232 galaxies per bin in order to match the total number of galaxies in the Galazy Zoo Biased (GZB) sample \cite{Bamford2009}. Note that the machine learning classifications of \cite{Huertas2011} (HC11) use the F07 classifications for training the Support Vector Machine.}
\label{fig:bias}
\end{figure*}

We start with the smallest dataset F07 with 1843 spirals and ellipticals. We find the best finite-sampling bias levels for a maximum bin number of 8 in the intrinsic parameters and 2 in the observed parameters, obtaining 115 galaxies per bin. Note that this is  the minimum bin sizes we can apply due to the number of intrinsic and biasing (observed in this case) properties in the data. Recall the data from F07 and NA10 only contains galaxies for $m < 16$, so again we consider galaxies with $m < 16$ in the GZ and HC11 datasets. Figure \ref{fig:bias}a shows the biases under these assumptions for different datasets. Again, due to the size of the standard deviation error bars of $L$, it is hard to make statistical significant conclusions on the difference between datasets. 

If we exclude F07, we find a maximum binning size of $2^5$ in the intrinsic parameters  and 3 bins in the observable parameters with 144 galaxies per bin. Figure \ref{fig:bias_Bamford}b shows the results of the bias for each data set. HC11 presents the lowest bias. Expert labels froms NA10 are less biased than Galaxy Zoo. With this number of galaxies there is no statistical significance between GZB and GZD for a given probability threshold.

\long\def\/*#1*/{}

If we exclude both F07 and NA10, we are able to consider galaxies with $m >16$ and use $2^8$ bins for the intrinsic parameters and 4 bins for the biasing parameter $r_\mathrm{PSF}$, from which we obtain 232 objects per bin. 
In Figure \ref{fig:bias}c we show the labeling bias as defined by Equation \ref{eq:L} using this binning strategy for GZB, GZD, and HC11. The highest values of $L$ are obtained over the GZ biased data sets and the lowest labeling bias is obtained for HC11. With this amount of data we notice that GZD with $p>0.5$ shows a smaller amount of bias than GZB. At the same time, by choosing $p>0.8$ the selected GZD data set is significantly more biased than the GZD data with $p>0.5$ and closer to GZB for $p>0.5$. In other words, it appears that the de-biasing procedure implemented in \cite{Bamford2009} for Galaxy Zoo classifications does not work when the vast majority of classifiers agree on the morphological type.

\begin{figure}
\centering
\includegraphics[width =0.45 \textwidth] {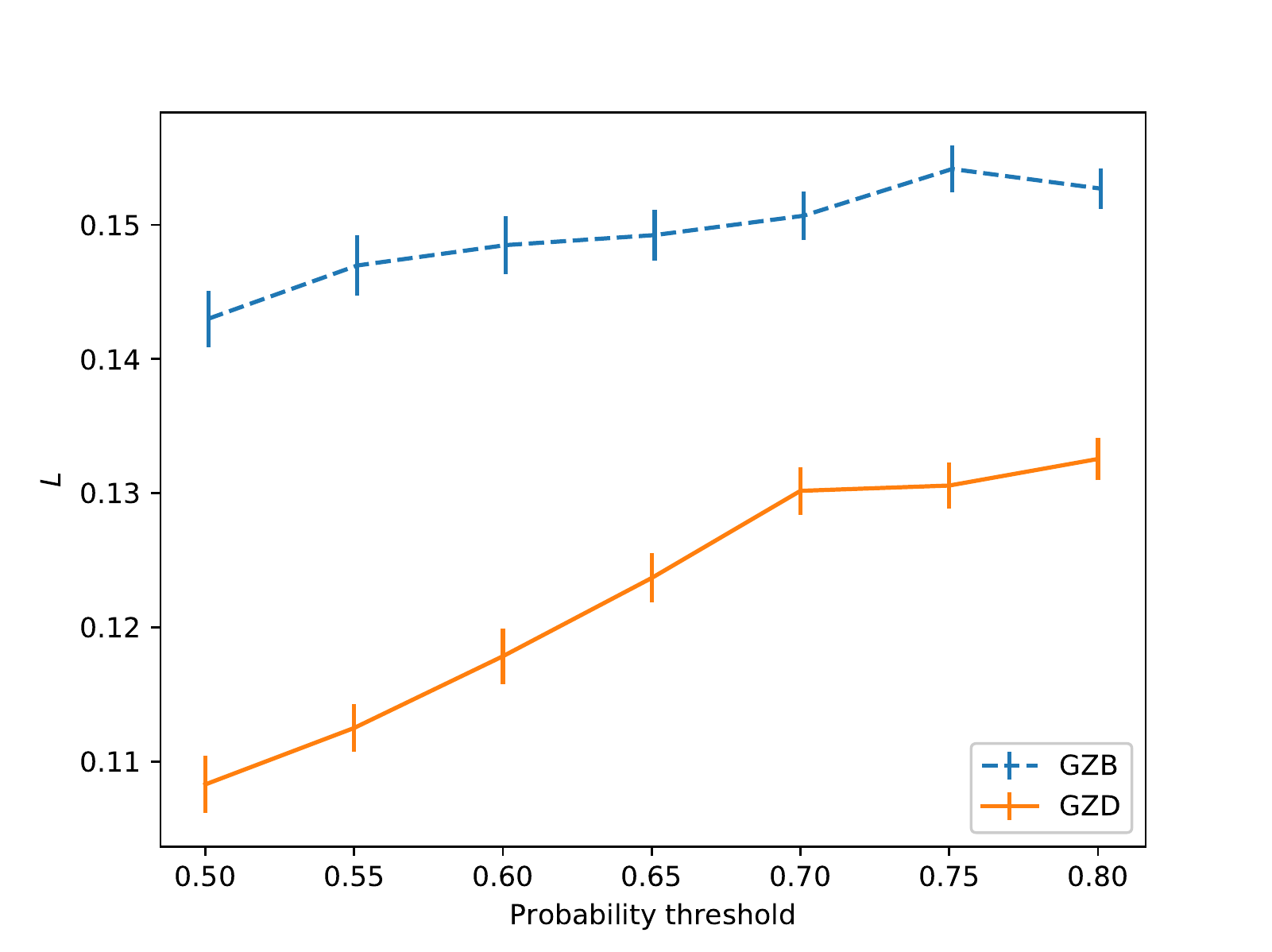}
\caption{Bias $L$ for the Galaxy Zoo Biased sample (GZB) and the Galaxy Zoo Debiased sample (GZD) versus the probability threshold used to define the classes. Notice that the GZB bias does not significantly decrease with increasing probability threshold and that the GZD bias increases with increasing probability threshold. An explanation for these unexpected trends is discussed in the text. }
\label{fig:L_vs_pbb}
\end{figure}

We explore this interesting result further in Figure \ref{fig:L_vs_pbb}, where we plot the bias $L$ as a function of an increasing Galaxy Zoo classification probability threshold. For the biased sample, we see no clear trend. However, the de-biased sample shows a trend of increasing bias with increasing classification probability threshold. 

We can explain Figure  \ref{fig:L_vs_pbb} in the following way. First, \cite{Bamford2009} use a statistical correction (their equations A3 and A4) that depends on both the raw classification probabilities as well as the intrinsic characteristics of the galaxy (e.g., absolute magnitude, physical Petrosian radius, redshift). This form of correction was chosen under the assumption that at high classification probabilities, no morphology adjustment should be applied since the labels would be correct (see Figure A9 of \cite{Bamford2009}). Thus, the fact that the classification bias is closer to the one from GZB at high $p$ in Figure \ref{fig:L_vs_pbb} 
stems from the design of the classification adjustment formalism. Since the correction term approaches zero at high $p$, the sample reverts back to the same level of bias inherent in the nominal biased sample. 

What is perhaps surprising is that for the Galaxy Zoo Biased (GZB) sample, the level of bias does not decrease as the classifications reach higher levels of confidence (high $p$). Recall from the Introduction that our algorithm aims to quantify the presence of classification bias due to mislabeled data. We noted that such mislabeling error is not a statistical labeling error, but instead an intrinsic error related to the quality of data itself (see Figure \ref{fig:S_gals}). The high bias at $p = 0.5$ in the Galaxy Zoo Biased sample is to be expected, especially for spirals, when the data quality is low or when the classifiers are non-expert. As noted earlier, it can be difficult to distinguish between spirals and ellipticals due to the data quality at low brightness or small apparent size. At $p > 0.8$, it should have been easy for classifiers to have identified morphologies since the classifications from different classifiers agree. This is likely to be true for spirals, but it is nearly impossible for the classifiers to separate ellipticals from spirals when the data quality is bad. When the data is bad the classifications will always tend towards elliptical with high confidence. In other words, while it is almost certainly the case that $p>0.8$ spiral galaxies are true spirals, $p>0.8$ ellipticals are not always true ellipticals. Thus the formalism to adjust classifications for ellipticals should not converge to the raw classification, even at high $p$.

An alternative approach to correct biased labels is to produce a set of simulated calibration images. These images are degraded versions of high quality images, where the ground truth labels can be accurately estimated. Galaxy Zoo: Hubble \citep{Willett2017} label such images through their interface, producing a set of biased labels with their corresponding ground truth labels. Their correction term allows high classification probabilities to be adjusted. Measuring biases on such corrected labels would be very interesting, but slightly out of the scope of this paper: here, we present a metric to assess biases, and show an application to low redshift galaxies. We plan to address biases at higher redshift galaxies in the future, including Galaxy Zoo: Hubble.

The final question regarding Figure \ref{fig:bias} is why the machine learning algorithms perform better than the training sets they used? Under perfect conditions the learned classifications should recover any biases inherent to the input training sets. Recall that HC11 uses an SVM supervised machine learning algorithm that is trained on the F07 data set. However, since the bias in the F07 data is higher than in the HC11 data, we conclude that the supervised machine learning technique used by \cite{Huertas2011} was able to mitigate the biases inherent in their training sets. 

It is important to recognize that labeling bias mitigation can only
occur if the ``correct'' choice of observed features are used in the
machine learning training sets. The term ``correct'' simply means that
the chosen observable parameters can in fact cause bias and that this
bias can be removed with additional truth information. In other words,
the bias caused by the apparent sizes can be fixed by leveraging
information about the true sizes and absolute magnitudes. As a counter
example, one should be able to feed the SVM-tool with an n-dimension
set of observed parameters that precisely recover the training set
classification (i.e., 100\% accuracy to the original training set). In
this case, one would have the same level of bias in the SVM trained
classifications as in the eyeballed training set. In the particular
case of HC11, they used features such as colors, shape and
concentration. These features correlate with morphological types
independent of observational parameters, such as
resolution. Therefore, the chosen HC11 observed galaxy parameter set
is enabling the Fukugita classification biases to be minimized by the
complexities of the machine learning algorithm. At the same time,
  a machine learning model trained over features such as colors, will
  not be able to correctly identify morphologies of outlier galaxies,
  such as red spirals or blue ellipticals. These galaxies may still be
  recognized by a human from a relatively high quality image. Machine
  learning models and eyeball labels may be complementarily used to
  obtain scientifically interesting outliers.

%
%


\newlength\replength
\newcommand\repfrac{.33}
\newcommand\dashfrac[1]{\renewcommand\repfrac{#1}}
\setlength\replength{1.5pt}
\newcommand\rulewidth{.6pt}
\newcommand\tdashfill[1][\repfrac]{\cleaders\hbox to \replength{%
  \smash{\rule[\arraystretch\ht\strutbox]{\repfrac\replength}{\rulewidth}}}\hfill}
\newcommand\tabdashline{%
  \makebox[0pt][r]{\makebox[\tabcolsep]{\tdashfill\hfil}}\tdashfill\hfil%
  \makebox[0pt][l]{\makebox[\tabcolsep]{\tdashfill\hfil}}%
  \\[-\arraystretch\dimexpr\ht\strutbox+\dp\strutbox\relax]%
}

\section {Conclusions}\label{sec:Conclusions}

Observational parameters, such as resolution, can bias the procedure of human labeling of galaxies. We have developed a metric to assess systematic mislabeling of galaxy morphologies which incorporates information about the galaxy intrinsic parameters, such as their true sizes and absolute magnitudes. Our algorithm requires that the true (but unknown) fractions for the classes be constant when binned against their intrinsic parameters. We then quantify the mean deviation of the fraction of objects from the estimated intrinsic fraction in terms of their observational parameters.

We then conduct a relative comparison of labeling bias for expert, citizen science, and machine learning-based galaxy classifications between spirals and ellipticals (+S0s). We find that, when enough data is provided, the bias in expert labels is statistically lower than the citizen science labels. We use our metric to recover Galaxy Zoo de-biasing procedure, under the assumption that labels are biased in terms of the redshift. By using the labeled image resolution as biasing parameters instead, we show our metric is able to find biases that have not been addressed. These biases may be statistically corrected in the future in the same manner that Galaxy Zoo does it. The classifications which use machine learning techniques show the least levels of bias, even when they are trained on biased ``gold standards''. We conclude that future large-scale morphological classification efforts should employ a combination of human classifications and machine learning in order to minimize labeling bias.

In this paper we have focused on the problem of galaxy morphologies. However, our approach may be applied to any other labeled data set where intrinsic information can be inferred. We have made our code publicly available so that it can be used by the galaxy evolution community or any other classification problem at https://github.com/guille-c/labeling\_bias.\footnote{Licensed under the terms of the GNU General Public License v3.0.}.

\section*{Acknowledgements}
We wish to thank Nancy Hitschfeld, Benjam\'in Bustos, Eduardo Vera,
Jaime San Mart\'in, Chris Smith, and Alfredo Zenteno for valuable discussion and
supporting our project.

G.C.V. gratefully acknowledge financial support from CONICYT-Chile
through its FONDECYT postdoctoral grant number
3160747; CONICYT-Chile and NSF through the Programme of
International Cooperation project DPI201400090; Basal Project PFB--03; the
Ministry of Economy, Development, and Tourism's Millennium Science
Initiative through grant IC120009, awarded to The Millennium Institute
of Astrophysics (MAS). CJM was supported by the National Science Foundation under Grant No. 1256260. 
Powered\@NLHPC: This research was partially supported by the
supercomputing infrastructure of the NLHPC (ECM-02). Most of the table
operations and plots were done using TOPCAT \citep{Taylor2005} and
matplotlib \citep{Hunter2007}. We used numpy
  \citep{Oliphant2006}, scipy \citep{Oliphant2007}, and pandas
  \citep{McKinney2010} for numerical computations. The kernel density
  estimation model was trained using scikit-learn
  \citep{scikit-learn}.
  
  Funding for the SDSS and SDSS-II has been provided by the Alfred
  P. Sloan Foundation, the Participating Institutions, the National
  Science Foundation, the U.S. Department of Energy, the National
  Aeronautics and Space Administration, the Japanese Monbukagakusho,
  the Max Planck Society, and the Higher Education Funding Council for
  England. The SDSS Web Site is http://www.sdss.org/.
  
  The SDSS is managed by the Astrophysical Research Consortium for the
  Participating Institutions. The Participating Institutions are the
  American Museum of Natural History, Astrophysical Institute Potsdam,
  University of Basel, University of Cambridge, Case Western Reserve
  University, University of Chicago, Drexel University, Fermilab, the
  Institute for Advanced Study, the Japan Participation Group, Johns
  Hopkins University, the Joint Institute for Nuclear Astrophysics,
  the Kavli Institute for Particle Astrophysics and Cosmology, the
  Korean Scientist Group, the Chinese Academy of Sciences (LAMOST),
  Los Alamos National Laboratory, the Max-Planck-Institute for
  Astronomy (MPIA), the Max-Planck-Institute for Astrophysics (MPA),
  New Mexico State University, Ohio State University, University of
  Pittsburgh, University of Portsmouth, Princeton University, the
  United States Naval Observatory, and the University of Washington.

Based on observations made with the NASA/ESA Hubble Space Telescope, and obtained from the Hubble Legacy Archive, which is a collaboration between the Space Telescope Science Institute (STScI/NASA), the Space Telescope European Coordinating Facility (ST-ECF/ESA) and the Canadian Astronomy Data Centre (CADC/NRC/CSA).


\bibliographystyle{apj}
\bibliography{ms}{}

\end{document}